\documentclass[aps,superscriptaddress,showpacs,nofootinbib,floatfix,epfs]{revtex4}
\pdfoutput=1
\usepackage{amsfonts,amscd,amsmath, amssymb,graphicx,color}

\def\W{W_{\text{\tiny gen}}}

\def\f0{f_{\text{\tiny 0}}}
\newcommand{\oh}{\frac{1}{2}}
\def\z{z_{\text{\tiny 0}}}
\def\ep{\text{e}}

\def\g{\mathfrak{g}}
\def\oh{\frac{1}{2}}
\def\x0{x_{\text{\tiny 0}}}
\def\y0{y_{\text{\tiny 0}}}
\def\r0{r_{\text{\tiny 0}}}
\def\z0{z_{\text{\tiny 0}}}
\def\m{\mathfrak{m}}
\def\s{\mathfrak{s}}
\def\last{\lambda_\ast}
\def\h{\text{e}^{\s r^2}}

\begin{document}
\title{Exotic Hybrid Quark Potentials}
\author{Oleg Andreev}
 \affiliation{L.D. Landau Institute for Theoretical Physics, Kosygina 2, 119334 Moscow, Russia}
\affiliation{Arnold Sommerfeld Center for Theoretical Physics, LMU-M\"unchen, Theresienstrasse 37, 80333 M\"unchen, Germany}
\begin{abstract} 
We use gauge/string duality to model some hybrid heavy-quark potentials. The potentials under consideration can't be described by a single Nambu-Goto string. This is why we call them "exotic". For $\Sigma_u^-$, the result is in quite good agreement with lattice simulations.
\end{abstract}
\pacs{12.39.Pn, 12.90.+b, 12.38.Lg}
\preprint{LMU-ASC 46/12}
\maketitle

\vspace{-7.5cm}
\begin{flushright}
LMU-ASC 46/12
\end{flushright}
\vspace{6cm}

\section{Introduction}
Static quark potentials play an important role in the spectrum and phenomenology of both heavy mesons and heavy baryons. In many cases they 
have been computed in lattice simulations, and the results reveal a remarkable agreement with phenomenology.\footnote{For a review, see \cite{bali}.}

One of the implications of the AdS/CFT correspondence \cite{malda} is that it became a new tool to deal with strongly coupled gauge theories 
and, as a result, resumed interest in finding a string description of strong interactions. 

In this paper we continue a series of studies \cite{az1,abar} devoted to the heavy-quark potentials within a five(ten)-dimensional effective string theory. The model is not exactly dual to QCD, but it may be useful as a guide in the present period where the string dual to QCD is unavailable. 
In \cite{az1}, the quark-antiquark potential was computed. The resulting potential is Coulomb-like at short distances and linear at long range. 
Subsequent work \cite{white} has made it clear that the model should be taken seriously, particular in the context of consistency with the available 
lattice data as well as phenomenology. In \cite{abar}, the static three-quark potential and pseudo-potential were studied. The results support the Y-ansatz for the baryonic area law as well as the universality of the string tension.

The question naturally arises: What happens when the model is used for computing hybrid potentials? These potentials have been the object of many studies \cite{bali}, particularly in relation to hybrid mesons and baryons \cite{isgur}. Importantly, (exotic) hybrid mesons are one of the goals of the GlueX experiment planned at Jefferson Lab.

We present here an example of the hybrid potentials derived from the effective string theory in higher dimensions. It contains the 
quark-antiquark potential of \cite{az1} in an appropriate limit, but it includes as well a new object called "defect". A defect is nothing but a macroscopic description of some gluonic degree of freedom in strong coupling. 

The paper is organized as follows. In section II, we discuss the five(ten)-dimensional effective string theory. We begin by summarizing the theoretical 
background and the results of \cite{az1} for the quark-antiquark potential. Then, we introduce a point-like defect to macroscopically 
describe a class of string excitations and, as a result, some hybrid potentials. We go on in Sec.III to discuss the lattice results as well as phenomenology. We conclude in Sec.IV with a brief discussion of possibilities for further study.

\section{Some Hybrid Quark Potentials via Gauge/string Duality} 
\renewcommand{\theequation}{2.\arabic{equation}}
\setcounter{equation}{0}
\subsection{General formalism}

First, let us set the basic framework. The background metric in question is a one-parameter deformation of the Euclidean 
$\text{AdS}_5$ space of radius ${\cal R}$ \cite{az1}

\begin{equation}\label{metric}
ds^2={\cal R}^2 w \bigl(dt^2+d\vec x^2+dr^2\bigr)\,,
\quad w(r)=\frac{\h}{r^2}
\,,
\end{equation}
where $d\vec x^2=dx^2+dy^2+dz^2$. $\s$ is a deformation parameter. In addition, we take a constant dilaton and discard other background fields. 

As a prelude to discussing the hybrid potentials, let us briefly consider the quark-antiquark potential (the ground state energy of flux tube). 
The potential can be determined from the expectation value of a Wilson loop. We do so by adopting the proposal 
of \cite{malda1}

\begin{equation}\label{wilson}
	\langle W(C)\rangle\sim \ep^{-S}
	\,,
\end{equation}
where $S$ is an area of a string worldsheet bounded by a curve $C$ at the boundary of the deformed AdS space.

As known, if $C$ is a rectangular loop of size $R\times T$, then $\langle W(C)\rangle\sim\ep^{-V(R)T}$ as $T\rightarrow\infty$. For the 
background geometry \eqref{metric}, the potential $V(R)$ is written in parametric form as \cite{az1}

\begin{equation}\label{r}
R(\lambda)=2\sqrt{\frac{\lambda}{\s}}
\int_0^1 dv\,v^2\ep^{\lambda(1-v^2)}
\biggl(1-v^4\ep^{2\lambda(1-v^2)}\biggr)^{-\oh}
\,,
\end{equation}

\begin{equation}\label{v0}
V(\lambda)=2\g\sqrt{\frac{\s}{\lambda}}
\int_0^1 \frac{dv}{v^2}\Biggl[\ep^{\lambda v^2}
\biggl(1-v^4\ep^{2\lambda (1-v^2)}\biggr)^{-\oh}-1-v^2\Biggr] + C
\,,
\end{equation}
where $\lambda$ is a parameter taking values in $[0,1]$, $C$ is a normalization constant, and $\g=\frac{{\cal R}^2}{2\pi\alpha'}$.

The potential is linear at long distances

\begin{equation}\label{lr}
	V(R)=\sigma R +C+o(1)
	\,,
\end{equation}
where $\sigma=e\g\s$ is a string tension. While it is Coulomb-like at short distances

\begin{equation}\label{sr}
	V(R)=-\frac{\alpha}{R}+C+o(1)
	\,,
\end{equation}
where $\alpha=(2\pi)^3\g/\Gamma^4(\frac{1}{4})$.

\subsection{Exotic hybrids}

The quark-antiquark potential corresponds to the ground state energy of a single string stretched between two fermionic sources. The common wisdom is 
that excited strings (fluxes) lead to hybrid potentials \cite{isgur}. The structure of string excitations is quite complicated and is made up of 
many different kinds of elementary excitations like vibrational modes, loops of flux, knots, kinks, disconnected loops etc. 

Adopting the view point that some excitations remain very narrow along a string, we will model them by inserting local objects 
on a string. We will call these objects defects. The reason for this is that in the presence of defects string embeddings of worldsheets into 
spacetime are not differentiable at points where defects are located. In this case any use of the Nambu-Goto action alone to describe string excitations is no longer appropriate. Certainly, not all of excitations may be described by our construction. So we are bound to learn something if we succeed.
 
In the case of a single defect, our construction is quite similar to that of \cite{abar} for the multi-quark potentials. So, we place a quark-antiquark pair at the boundary points of 
the five(ten)-dimensional space and consider a configuration in which each fermionic source is the endpoint of a fundamental string. The strings join at 
a defect in the interior as shown in Fig.1.
\begin{figure}[htbp]
\centering
\includegraphics[width=5.75cm]{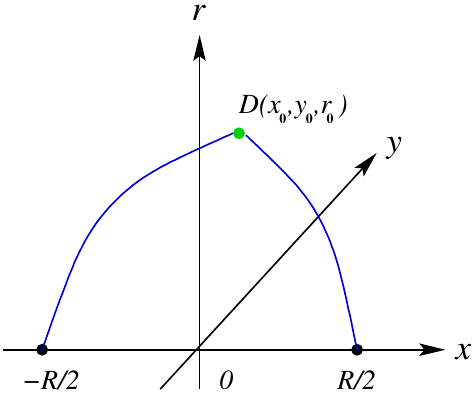}
\caption{{\small A configuration used to calculate the hybrid potentials. The quark and antiquark are set at $x=R/2$ and $x=-R/2$, respectively. The defect is placed at $D$.}}
\end{figure}
Thus, the action has in addition to the standard Nambu-Goto actions, also a contribution arising 
from the defect. It is given by  

\begin{equation}\label{stringaction}
S=\sum_{i=1}^2  S_i +S_{\text{\tiny def}}
\,,
\end{equation}
where $S_i$ denotes the action of the string connecting the $i$-source with the defect.

In the lattice formulation the hybrid potentials are extracted from generalized Wilson loops in the limit $T\rightarrow\infty$ \cite{bali}. In our framework, a natural proposal for the expectation value of the generalized Wilson loop is 

\begin{equation}\label{wloop-ads}
\langle\,\W \,\rangle\sim \ep^{-S_{\text{\tiny min}}}
\,,
\end{equation}
where $S_{\text{\tiny min}}$ is an area of two string worldsheets bounded by a loop $C$ at the boundary of AdS space and glued along a line in 
the bulk. As in \cite{az1}, in the large $T$ limit $S_{\text{\tiny min}}=T\tilde V$ such that a dominant contribution comes from a lateral surface
whose area is proportional to $T$, while two end-surfaces only provide small corrections.

Since we are interested in a static configuration, for the strings we take 
 
\begin{equation}\label{gauge}
t_i(\tau_i)=\tau_i\,,\quad
x_i(\sigma_i)=a_i\sigma_i+b_i
\,.
\end{equation}
Here $(\tau_i,\sigma_i)$ are worldsheet coordinates. The Nambu-Goto action of the $i$-string is then

\begin{equation}\label{ng-action}
S_i=T\g\int_0^1 d\sigma_i\,w\sqrt{a^2_i+ y'^2_i+r'^2_i}
\,,
\end{equation}
where a prime denotes a derivative with respect to $\sigma_i$.

The action for the defect is taken to be of the form

\begin{equation}\label{vertex}
S_{\text{\tiny def.}}=T{\cal V}(\r0 )
\,,
\end{equation}
where ${\cal V}$ can be considered as its effective potential. Unfortunately, the explicit form of ${\cal V}$ is not determined only 
from the 5-dimensional metric. It requires the knowledge of the string theory dual to QCD. We will return to this issue in the 
next sections.

The boundary conditions on the fields are given by 

\begin{align}\label{boundary}
x_1(0)&=-R/2\,, 
&
x_1(1)&=\x0\,,
&
y_1(0)&=0\,,
&
y_1(1)&=\y0\,,
&
r_1(0)&=0\,,
&
r_1(1)&=\r0\,,\\
x_2(0)&=\x0\,, 
&
x_2(1)&=R/2\,,
&
y_2(0)&=\y0\,,
&
y_2(1)&=0\,,
&
r_2(0)&=\r0\,,
&
r_2(1)&=0\,.
\end{align}
These determine the coefficients $a_i$  and $b_i$ in \eqref{gauge}. Thus, we have

\begin{equation}\label{y}
a_1=\x0+R/2\,,\quad b_1=-R/2\,,
\quad
a_2=R/2-\x0\,,
\quad
b_2=\x0
\,.
\end{equation}

Now, we extremize the total action $S$ with respect to the worldsheet fields $y_i(\sigma_i)$ and 
$r_i(\sigma_i)$ describing the strings as well as with respect to $\x0$, $\y0$ and $\r0$ describing the location of the defect, with the 
following identifications: $\delta y_1(1)=\delta y_2(0)=\delta\y0$ and $\delta r_1(1)=\delta r_2(0)=\delta\r0$. In doing so, we use the fact that 
there are two symmetries which simplify the further analysis.

Since the integrand in \eqref{ng-action} does not depend explicitly on $\sigma_i$, we get the first integral of Euler-Lagrange equations

\begin{equation}\label{1integral}
I_i=\frac{w_i}{\sqrt{a_i^2+y_i'^2+r_i'^2}}
\,.
\end{equation}
In addition, because of translational invariance along the $y$-direction, there is another first integral. Combining 
it with \eqref{1integral} gives

\begin{equation}\label{2integral}
P_i=y'_i\,.
\end{equation}
Together with the boundary conditions these equations determine the $y_i$'s

\begin{equation}\label{x}
y_1 (\sigma_1 )=\y0\sigma_1\,,\quad
y_2(\sigma_2)=-\y0\sigma_2+\y0
\,.
\end{equation}

Next, we extremize the action with respect to the location of the defect. After using \eqref{y} and \eqref{x}, we get 

\begin{equation}\label{xyr-eqs}
\left(\x0+R/2\right)I_1+\left(\x0-R/2\right)I_2=0
\,,
\quad
\y0\left(I_1+I_2\right)=0
\,,
\quad
r_1'(1)I_1-r_2'(0)I_2+\g^{-1}{\cal V}'(\r0)=0
\,.
\end{equation}
From this it follows that $\y0=0$. As a result, the static configuration lies in the $xr$-plane.

Now we introduce $k_1=\left(\frac{r_1'(1)}{\x0+R/2}\right)^2$ and $k_2=\left(\frac{r_2'(0)}{R/2-\x0}\right)^2$. This allows us to rewrite 
the first equation of \eqref{xyr-eqs} as 

\begin{equation}\label{xr}
	\frac{1}{\sqrt{1+k_1}}-\frac{1}{\sqrt{1+k_2}}=0\,.
	\end{equation}
Obviously, it yields the unique solution $k_1=k_2=k$. Combining this with \eqref{xyr-eqs}, we have

\begin{equation}\label{r0}
	\frac{2}{\sqrt{1+k^{-1}}}+\frac{1}{\g}\frac{\cal V'}{w}(\r0)=0
	\,.
\end{equation}

If we define the first integrals $I_1$ at $\sigma_1=1$ and $I_2$ at $\sigma_2=0$ and integrate over $[0,1]$ of 
$d\sigma_i$, then by virtue of \eqref{xr} we get

\begin{equation}\label{l-eqs}
R\pm 2\x0=2
\sqrt{\frac{\lambda}{\s(1+k)}}
\int_0^1 dv\,v^2\ep^{\lambda(1-v^2)}
\biggl(1-\frac{1}{1+k}v^4\ep^{2\lambda(1-v^2)}\biggr)^{-\oh}
\,,
\end{equation}
where $\lambda=\s\r0^2$. Obviously, it is consistent only if $\x0=0$. As a result, we end up with the most symmetric configuration.

Now, we will compute the energy of the configuration. First, we reduce the integrals over $\sigma_i$ in Eq.\eqref{ng-action} to that 
over $r_i$. This is easily done by using the first integral \eqref{1integral}. Since the integral is divergent at $r_i=0$, we regularize it by imposing 
a  cutoff $\epsilon$.\footnote{Importantly, in this process we use the same renormalization scheme as that for $V$ in \cite{az1}.} 
Finally, the regularized expression takes the form

\begin{equation}\label{energy}
\tilde V_R=
{\cal V}(\lambda)+
2\g\sqrt{\frac{\s}{\lambda}}
\int_{\sqrt{\frac{\s}{\lambda}}\epsilon}^1 \frac{dv}{v^2}\, \ep^{\lambda v^2}
\biggl(1-\frac{1}{1+k}v^4\ep^{2\lambda(1-v^2)}\biggr)^{-\oh}
\,.
\end{equation}
Its $\epsilon$-expansion is 

\begin{equation*}\label{energy2}
\tilde V_R=\frac{2\g}{\epsilon}+O(1)
\,.
\end{equation*}
Subtracting the $\frac{1}{\epsilon}$ term (quark masses) and letting $\epsilon=0$, we find a finite result

\begin{equation}\label{energy3}
\tilde V(\lambda)={\cal V}(\lambda)
+2\g\sqrt{\frac{\s}{\lambda}}
\int_0^1 \frac{dv}{v^2}\Biggl[\ep^{\lambda v^2}
\biggl(1-\frac{1}{1+k}v^4\ep^{2\lambda (1-v^2)}\biggr)^{-\oh}-1-v^2\Biggr] + C
\,,
\end{equation}
where $C$ is the same normalization constant as in \eqref{v0}. Note that \eqref{energy3} reduces to \eqref{v0} at ${\cal V}=0$, as promised in the 
introduction.\footnote{It follows from \eqref{r0} that $k=0$ at ${\cal V}=0$.}

\subsection{Concrete examples}

We will now describe a couple of concrete examples in which one can develop a level of understanding that is somewhat similar to that of the 
quark-antiquark case \cite{az1}. 

\subsubsection{Model A}

First, we consider the five-dimensional geometry \eqref{metric}. In this case we specify the action $S_{\text{\tiny def}}$ as that of a 
point like particle $\m\int ds$, with $\m$ a particle mass. The latter implies that 

\begin{equation}\label{potential}
{\cal V}(\r0 )=\m{\cal R}\,\text{e}^{\oh\s\r0^2}/\r0
\,.
\end{equation}

Now \eqref{r0} takes the form

\begin{equation}\label{r-eqs}
\frac{1}{\sqrt{1+k^{-1}}}=\kappa\left(1-\lambda\right) \ep^{-\oh\lambda}
\,,
\end{equation}
where $\kappa=\frac{\m{\cal R}}{2\g}$.

With the form \eqref{r-eqs}, $k$ can be explicitly computed as a function of $\lambda$. Inserted in \eqref{l-eqs} and \eqref{energy3}, this gives

\begin{equation}\label{l-eqs-ex}
R(\lambda)=2\sqrt{\frac{\lambda}{\s}\rho}\int_0^1 dv\,v^2\ep^{\lambda (1-v^2)}\Bigl(1-\rho \,v^4\ep^{2\lambda (1-v^2)}\Bigr)^{-\oh}
\,
\end{equation}
and
\begin{equation}\label{energy-ex}
\tilde V(\lambda)=2\g\sqrt{\frac{\s}{\lambda}}
\biggl[\kappa\,\ep^{\oh\lambda}+
\int_0^1 \frac{dv}{v^2}\biggl(\ep^{\lambda v^2}
\Bigl(1-\rho \,v^4\ep^{2\lambda (1-v^2)}\Bigr)^{-\oh}-1-v^2\biggr)\biggr]+C
\,,
\end{equation}
where $\rho(\lambda)=1-\kappa^2\left(1-\lambda\right)^2 \ep^{-\lambda}$. A simple analysis shows that the 
parameter $\lambda$ takes values in the interval $[0,1]$ if $\kappa\leq 1$ and $[\last,1]$ if $\kappa>1$. Here $\last$ is a solution 
of equation $\rho(\lambda)=0$.

The analysis of Eqs.\eqref{l-eqs-ex} and \eqref{energy-ex} in the two limiting cases, long and short distances, is formally similar to that 
of \cite{abar}. From the 5d perspective, the system is prevented from getting deeper into the $r$-direction than $1/\sqrt{\s}$ (the soft wall). 

At long distances, the potential $\tilde V(R)$ is linear for any value of $\kappa$

\begin{equation}\label{largerA}
	\tilde V(R)=\sigma R+C+\Delta+o(1)
\,,
\end{equation} 
with the same string tension as in \eqref{lr}. 

The point here is a finite gap between $\tilde V$ and $V$ 

\begin{equation}\label{gap}
\Delta=\lim_{R\to\infty}\tilde V(R)-V(R)\,,\quad\text{with}\quad
\Delta=2\kappa\sqrt{\g\sigma}
\,.
\end{equation}
This is opposite to what has been found in 4d effective string theories, where there is no finite energy gap at long distances. For example, 
it is decreasing with increasing distance as $\Delta\sim 1/R$ \cite{arvis,LW}. Note that the gap is due to the effective potential \eqref{potential} 
evaluated at $\lambda=1$. 

A more involved analysis shows three types of asymptotic behavior at short distances: 

For $0<\kappa<1$, the potential is Coulomb-like 

\begin{equation}\label{smallrA1}
	\tilde V(R)=-\frac{\tilde\alpha(\kappa)}{R}+O(1)
	\,,
\end{equation}
where 
\begin{equation}\label{function-alpha}
\tilde\alpha (\kappa)=\frac{4}{3}\g\sqrt{1-\kappa^2}\,{}_2F_1\Bigl[\frac{1}{2},\frac{3}{4};\frac{7}{4};1-\kappa^2\Bigr]
\Bigl({}_2F_1\Bigl[-\frac{1}{4},\frac{1}{2};\frac{3}{4};1-\kappa^2\Bigr]-\kappa\Bigr)
\,.
\end{equation}
Note that $\tilde\alpha$ is positive for all allowed values of $\kappa$.

For $\kappa=1$, the potential is given by  

\begin{equation}\label{smallrA2}
	\tilde V(R)=C+\gamma\sqrt{R}+O(R)
	\,,
\end{equation}
where $\gamma=2\g\sqrt[4]{12\s^3}$.

Finally, for larger values of $\kappa$, the potential is given by 

\begin{equation}\label{smallrA3}
	\tilde V(R)= \tilde V(\last)+A(\last) R^2+O(R^3)
	\,.
\end{equation}
Here $\tilde V(\last)$ is given by \eqref{energy-ex} at $\lambda=\last$. At this value of $\lambda$ the integral can be performed to yield

\begin{equation}\label{tV}
\tilde V(\last)=2\g\sqrt{\frac{\s}{\last}}\Bigl(\kappa\,\ep^{\oh\last}-\ep^{\last}+\sqrt{\pi\last}\,\text{erfi}(\sqrt{\last})\Bigr)+C
\,.\qquad
\end{equation}
The coefficient $A$ is given by 
\begin{equation}\label{coefA}
A(\last)=\g\,\s^{\frac{3}{2}}\Bigl(\sqrt{\pi}\,\text{erf}(\sqrt{\last})-2\sqrt{\last}\,\ep^{-\last}\Bigr)^{-1}
\,.
\end{equation}
Note that $A$ is positive for all $\last$ of interest.

\subsubsection{Model B}

The above example has been formulated in a fashion that interprets the defect as a particle in 5 dimensions. Alternatively, one can consider 
a defect living in ten-dimensional space. For example, a flux loop is interpreted as a pair of baryon-antibaryon vertices connected by 
fundamental strings. According to \cite{edw}, in ten-dimensional space it is nothing but a brane-antibrane pair. Since the branes in question are fivebranes, it seems natural to assume that $S_{\text{\tiny def}}\sim\tau_5\int d^6x\sqrt{g^{(6)}}$, with $\tau_5$ a brane tension.

For subsequent applications, we will need to know the 10-dimensional geometry. One possibility here is to take \cite{a-pis} 

\begin{equation}\label{metric10}
ds^2={\cal R}^2 w \bigl(dt^2+d\vec x^2+dr^2\bigr)+\text{e}^{-\s r^2}g_{ab}^{(5)}d\omega^a d\omega^b 
\,.
\end{equation}
This is a deformed product of $AdS_5$ and a 5-dimensional compact space (sphere) $X$ whose coordinates are $\omega^a$. The deformation is due to the 
$r$-dependent warp factor. 

Because the fivebranes wrap on $\mathbf{R}\times X$, with $\mathbf R$ along the t-axis in the deformed $\text{AdS}_5$, we have

\begin{equation}\label{potential10}
{\cal V}(\r0 )=\m{\cal R}\,\text{e}^{-2\s\r0^2}/\r0
\,.
\end{equation}
Note that $\m\sim\tau_5\int d^5\omega\sqrt{g^{(5)}}$.

We are now ready to reproduce the potential. We begin with equation \eqref{r0} which now takes the form 

\begin{equation}\label{r-eqs10}
\frac{1}{\sqrt{1+k^{-1}}}=\kappa\left(1+4\lambda\right) \ep^{-3\lambda}
\,,
\end{equation}
with $\kappa$ as in \eqref{r-eqs}. As before, $k$ can be explicitly computed as a function of $\lambda$. Inserted in \eqref{l-eqs} 
and \eqref{energy3}, this gives

\begin{equation}\label{l-eqs-ex10}
R(\lambda)=2\sqrt{\frac{\lambda}{\s}\bar\rho}\int_0^1 dv\,v^2\ep^{\lambda (1-v^2)}\Bigl(1-\bar\rho \,v^4\ep^{2\lambda (1-v^2)}\Bigr)^{-\oh}
\,
\end{equation}
and
\begin{equation}\label{energy-ex10}
\tilde V(\lambda)=2\g\sqrt{\frac{\s}{\lambda}}
\biggl[\kappa\,\ep^{-2\lambda}+
\int_0^1 \frac{dv}{v^2}\biggl(\ep^{\lambda v^2}
\Bigl(1-\bar\rho \,v^4\ep^{2\lambda (1-v^2)}\Bigr)^{-\oh}-1-v^2\biggr)\biggr]+C
\,,
\end{equation}
where $\bar\rho(\lambda)=1-\kappa^2\left(1+4\lambda\right)^2 \ep^{-6\lambda}$. Now the parameter $\lambda$ takes values in the interval $[0,\lambda_c]$ if 
$\kappa<3/4\,\ep^{1/4}$, $[0,\lambda_{\star}]\cup[\last,\lambda_c]$ if $3/4\,\ep^{1/4}\leq\kappa<1$, and $[\last,\lambda_c]$ if $\kappa\geq1$. Here $\lambda_{\star}$ and $\last$ are solutions of equation $\bar\rho(\lambda)=0$ and $\lambda_c$ is a solution of equation $\bar\rho(\lambda)=\lambda^2\ep^{2(1-\lambda)}$. Together with \eqref{l-eqs-ex} and \eqref{energy-ex}, this is our main result.

The above derivation has been presented in a sketchy fashion. The basic idea is to consider a defect as a loop of flux. Such a defect being 
projected onto the boundary of ten-dimensional space is tiny but it spreads well in the internal space $X$. In general, the form of 
${\cal V}$ can be strongly dependent of $X$. In the situation considered above, we took ad hoc ten-dimensional geometry whose use is only supported by 
phenomenological applications \cite{a-pis}. With this choice of the metric, the potential \eqref{potential10} is decreasing with $r$. As a result, the system is allowed to go beyond the soft wall in the bulk. This makes it very different from 
single strings \cite{az1}, baryon vertices \cite{abar} and Model A, where it is not allowed to do so.
 
Now let us analyze Model B in more detail. As in \cite{az1}, it is unclear to us how to eliminate the parameter $\lambda$ and find $\tilde V$ as a 
function of $r$. We can, however, gain some important insights into the problem at hand from the two limiting cases.

The potential of Model B also shows a linear behavior at large distances 

\begin{equation}\label{largerB}
	\tilde V(R)=\sigma R +C+\Delta+o(1)
	\,,
\end{equation}
with the same string tension as in \eqref{lr}. It may look counter intuitive, but the point is that only a finite piece of the infinitely long 
string goes beyond the wall. This piece is not able to change the leading behavior at large $R$. But what it can do is modify the size of next-to-leading order corrections. The finite gap between $\tilde V$ and $V$ (at large $R$) is now 

\begin{equation}\label{gapB}
	\Delta= 2\sqrt{\ep\g\sigma}\biggl(\frac{\kappa}{\sqrt{\lambda_c}}\ep^{-1-2\lambda_c}+
	\int_1^{\sqrt{\lambda_c}}dv\sqrt{v^{-4}\ep^{2(v^2-1)}-1}\biggr)
	\,.
\end{equation}
Like in \eqref{gap}, the first term is a contribution of the local defect. It is nothing but a value of the effective potential \eqref{potential10}
at $\lambda=\lambda_c$. On the other hand, the second term represents a contribution of the string piece 
which gets beyond the wall.  

The short distance structure of the solution \eqref{l-eqs-ex10}-\eqref{energy-ex10} turns out to be richer than that of Model A. It also varies 
depending on a value of $\kappa$.

For $0<\kappa<3/4\,\ep^{1/4}$, the potential is Coulomb-like 

\begin{equation}\label{smallrB1}
	\tilde V(R)=-\frac{\tilde\alpha(\kappa)}{R}+O(1)
	\,,
\end{equation}
with $\tilde\alpha$ given by \eqref{function-alpha}.

For $3/4\,\ep^{1/4}\leq\kappa<1$, the solution \eqref{l-eqs-ex10}-\eqref{energy-ex10} in fact describes three different surfaces such that the expectation value of the generalized Wilson loop is given by $\langle\,\W \,\rangle=\sum_{i=1}^3w_i\,\ep^{-T\tilde V_i}$ with some weights $w_i.$ At short distances the $\tilde V$'s behave as

\begin{equation}
	\tilde V_i(R)=
	\begin{cases}
-\frac{\tilde\alpha (\kappa)}{R}+O(1)\,&\text{if}\,\,i=1\,,\\
\tilde V(\lambda_\star)+A(\lambda_\star)R^2+O(R^3)\,&\text{if}\,\,i=2\,,\\
\tilde V(\last)+A(\last)R^2+O(R^3)\,&\text{if}\,\,i=3\,.
\end{cases}
\end{equation}
Here $\tilde\alpha$ is given by \eqref{function-alpha}. $\tilde V(\lambda_\star)$ and $\tilde V(\last)$ are the values of the function \eqref{energy-ex10} at $\lambda=\{\lambda_\star,\last\}$. For these values of $\lambda$ the integral can be performed to yield 

\begin{equation}\label{functionB0}
\tilde V(x)=2\g\sqrt{\frac{\s}{x}}\Bigl(\kappa\,\ep^{-2x}-\ep^{x}+\sqrt{\pi x}\,\text{erfi}(\sqrt{x})\Bigr)+C
\,.
\end{equation}
Surprisingly, the coefficients in front of $R^2$ are expressed by the same function $A$ as that in Model A. In the large $T$ limit  $\tilde V_1$ is dominant at short distances and therefore provides $\tilde V(R)=\tilde V_1(R)$. Note that the surfaces corresponding to $\tilde V_1$ and $\tilde V_2$ exist at short distances only.

Finally, for larger values of $\kappa$, the potential behaves as 

\begin{equation}\label{smallB}
	\tilde V(R)=\tilde V(\last)+A(\last)R^2+O(R^3)
	\,,
\end{equation}
where $\tilde V(\last)$ and $A(\last)$ are given by \eqref{functionB0} and \eqref{coefA}, respectively.

\section{Phenomenological Implications}
\renewcommand{\theequation}{3.\arabic{equation}}
\setcounter{equation}{0}

Having derived the expressions for the potentials, we wish to address the issue of phenomenological implications. Otherwise, this 
will remain as an academic exercise in gauge/string duality.

\subsection{Comparison with lattice calculations}

The hybrid potentials are being extensively studied on the lattice \cite{bali}. They are classified by irreducible representations of 
the symmetry group $D_{4h}$. In practice, it is done by three quantum numbers. One of those 
is a projection of the total angular momentum on the axis passing through the quark and the antiquark. Our construction includes the system of 
two fundamental strings joined at the defect in the bulk such that its projection on the boundary entirely lies on the $x$-axis. Hence it has zero 
angular momentum. The latter means that we should look at the so called $\Sigma$ potentials. The two remaining quantum numbers correspond to 
symmetry (antisymmetry) under interchange of the string ends by a rotation by $\pi$ and by inversion in the midpoint. They are $(+,-)$ 
and $(g,u)$, respectively. In these notations the usual quark-antiquark potential $V$ is represented by $\Sigma_g^+$.

In this section, we will compare our analytic results with the data obtained in pure $SU(3)$ lattice gauge theory \cite{kuti}. For $V$ defined 
by \eqref{r} and \eqref{v0} the fitted parameters are $\g$, $\s$, and $C$. We fit their values with the potential $\Sigma_g^+$ of \cite{kuti}. So, 
we get $\g= 0.176$, $\s=0.44\,\text{GeV}^2$, and $C=0.71\,\text{GeV}$.\footnote{Interestingly, the fit from the slope of the $\rho$-meson 
Regge trajectories provides $\s=0.45\,\text{GeV}^2$ \cite{oa}. So, it seems that the value of $\s$ "slowly" depends on the flavor content.} With these parameters fixed, the potential $\tilde V$ has only a single free parameter $\kappa$. We then fit it with $\Sigma_u^-$ from \cite{kuti}. The results are plotted in Fig.2.  
\begin{figure}[htbp]
\centering
\includegraphics[width=5.25cm]{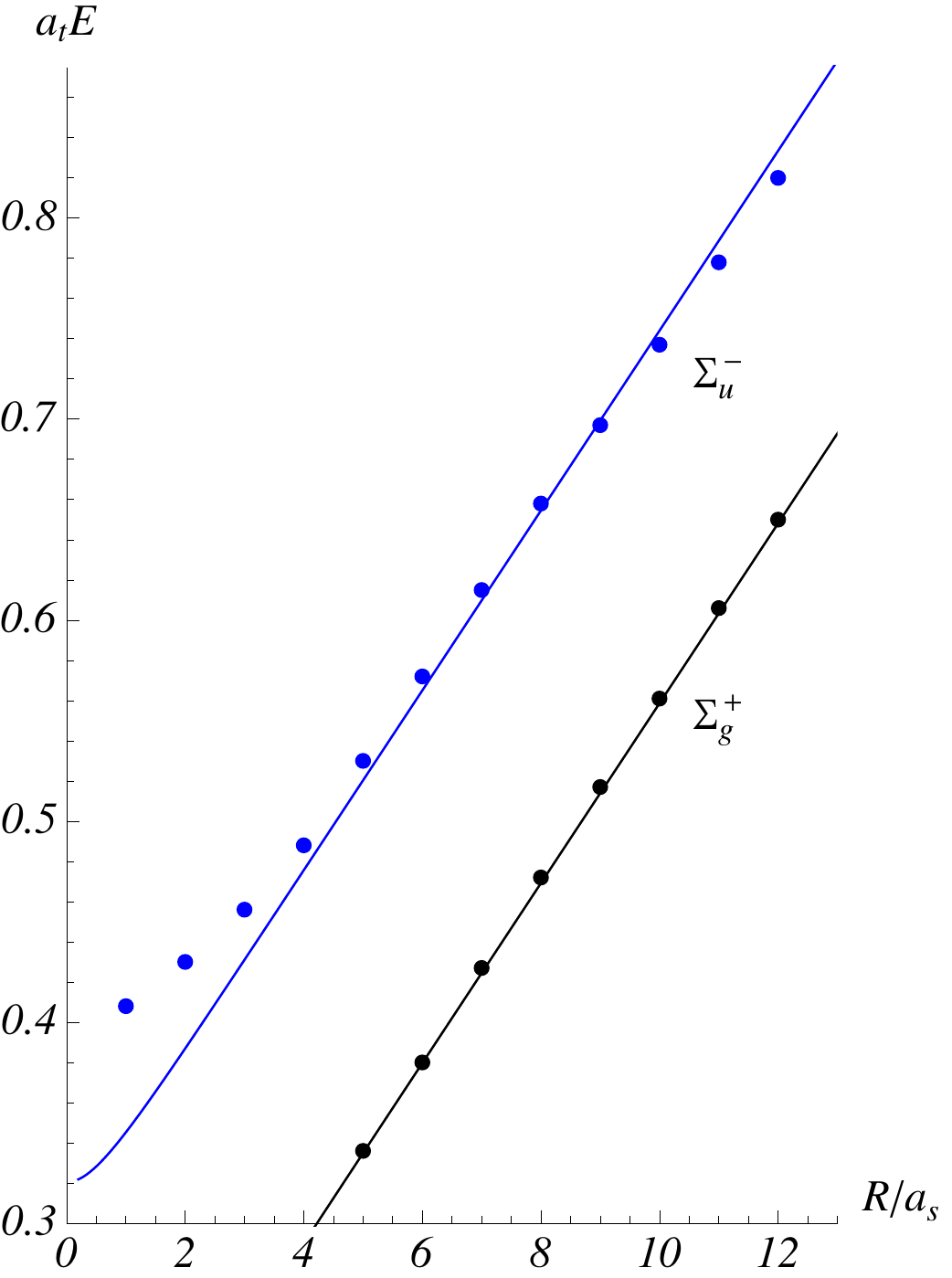}
\hspace{5cm}
\includegraphics[width=5.25cm]{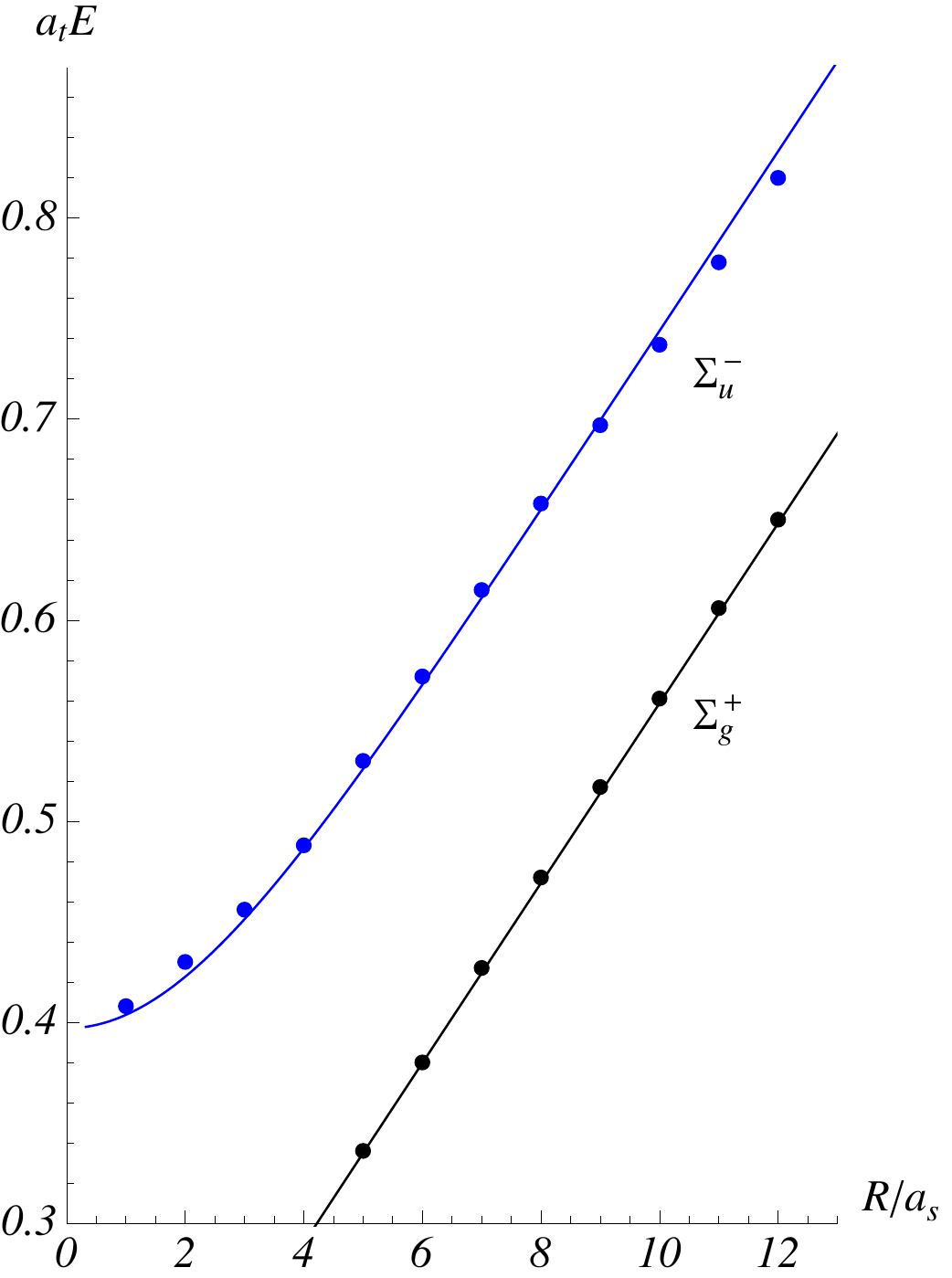}
\caption{{\small Energy levels $\Sigma_g^+$ and $\Sigma_u^-$. The dots represent the lattice data. The curves are our predictions. Left: Model A.
The value of $\kappa$ is set to $2.3$. Right: Model B. The value of $\kappa$ is set to $2000$. In both the cases we use the data and notations from Fig.2 of \cite{kuti}. So, we have $a_s=0.2\,\text{fm}$ and $a_t\approx 0.041\,\text{fm}$.}}
\end{figure}
We see that Model A shows a quite similar behavior to the lattice. However, for distances smaller than $0.6\,\text{fm}$ (3 in units of $a_s$) the deviations are large that makes the model not very attractive phenomenologically. Amazingly, Model B reproduces the lattice data for $\Sigma_u^-$ very well.  

\subsection{More discussions on the lattice and phenomenology}

One of the predictions of our model is a finite energy gap between the long string ground state and its $\Sigma_u^-$ excited state. Is this reasonable? No strong observational argument seems to exclude the possibility that at distances larger than $1.2\,\text{fm}$ the lattice data shown in Fig.2 are well fitted 
by separated parallel lines. On the other hand, one may reasonably argue that the range $1.2\,\text{fm}<R<2.4\,\text{fm}$ is too small for making 
such a conclusion.

Let us look at another portion of lattice data of \cite{kuti}. For $\Sigma_u^-$ and $\Sigma_g^-$ the energy gaps above the ground state are shown in Fig.3. 
\begin{figure}[htbp]
\centering
\includegraphics[width=6.5cm]{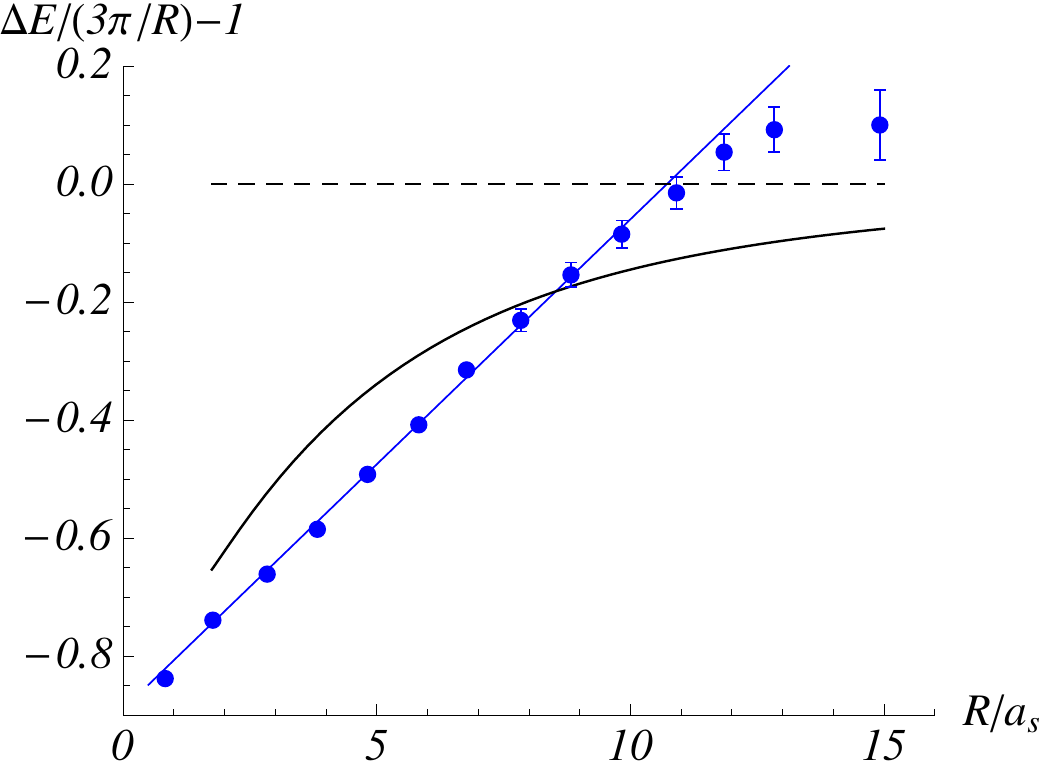}
\hfill
\includegraphics[width=6.5cm]{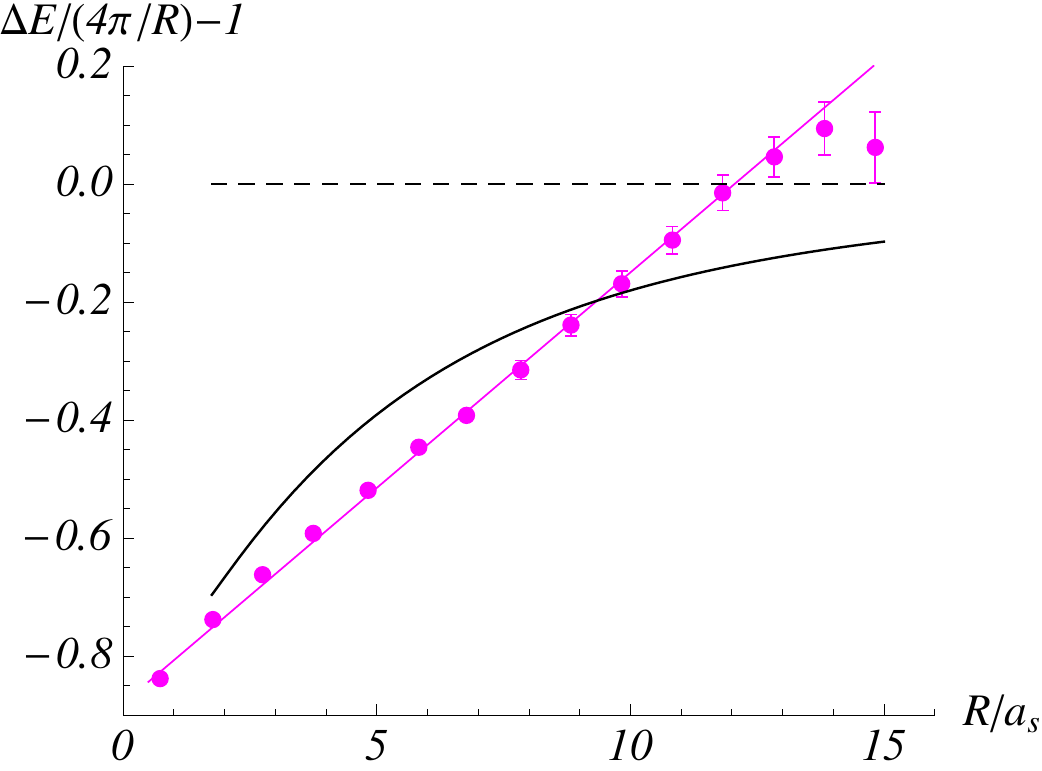}
\caption{{\small Energy gaps. The dots represent the lattice data. The black curves and lines are the predictions of the effective string models of \cite{LW} and \cite{arvis}. Left: The energy gap between $\Sigma_u^-$ and $\Sigma_g^+$. Right: The energy gap between $\Sigma_g^-$ and $\Sigma_g^+$. In both the cases we use the data and notations from Fig.1 of \cite{kuti}. So, we have $a_s=0.2\,\text{fm}$ and $a_t\approx 0.041\,\text{fm}$.}}
\end{figure}

For distances smaller than $2.5\,\text{fm}$ ($12.5$ in units of $a_s$), it seems reasonable to fit the data to the linear ansatz

\begin{equation}\label{la}
\Delta ER/(n\pi)-1=
\begin{cases}
0.083\,R/a_s-0.89\,&\text{if}\,\,n=3\,,\\
0.073 \,R/a_s-0.88\,&\text{if}\,\,n=4\,.
\end{cases}
\end{equation}
Here $n=3$ corresponds to $\Sigma_u^-$ and $n=4$ to $\Sigma_g^-$. Combining this with the successful parameterization of the ground state 
energy \footnote{It is successful if $R$ exceeds $0.5\,\text{fm}$. So, we have to drop only a couple of points from each plot shown in Fig.3, but 
that has no effect on our analysis.}

\begin{equation}\label{E0}
	E_0(R)=\sigma R+C-\frac{\pi}{12R}+o\Bigl(\frac{1}{R}\Bigr)
	\,
	\end{equation} 
gives

	\begin{equation}\label{fit}
		E_n(R)=\sigma R+C+\Delta_n+\frac{\pi}{R}\Bigl(Z_n n-\frac{1}{12}\Bigr)+o\Bigl(\frac{1}{R}\Bigr)
		\,,
	\end{equation}
where $Z_3\approx 0.11$, $Z_4\approx0.12$ and $\Delta_3\approx 0.77\,\text{GeV}$, $\Delta_4\approx 0.90\,\text{GeV}$. Note that 
the parameterization \eqref{fit} includes that of the ground state with $\Delta_0=0$.

To discuss effective string theories in a four-dimensional flat space, we must bear in mind the following: unlike those in 
a curved space (for discussion, see, e.g., \cite{af}), it is an attempt to describe long flux tubes. So, it makes sense in the large $R$ limit.  

The prediction of \cite{LW} for the string energy levels is given by 

\begin{equation}\label{LW}
		E_n(R)=\sigma R+C+\frac{\pi}{R}\Bigl(n-\frac{1}{12}\Bigr)+\dots
		\,,
		\end{equation}
where the dots denote higher order terms in $1/R$. For $n=0$, it reduces to \eqref{E0} that, as noted above, provides the successful parameterization of 
the ground state energy. However, it turns out to be less efficient for the excited states. The result is displayed in Fig.3 by the horizontal 
dashed lines. Obviously, for distances smaller than $2\,\text{fm}$ the lattice data are very far from the prediction \eqref{LW}.
		
A possible way out could be to calculate higher order corrections to \eqref{LW}. In \cite{arvis}, it was found

\begin{equation}\label{Arvis}
			E_n(R)=\sigma R\biggl[1+\frac{2\pi}{\sigma R^2}\Bigl(n-\frac{1}{12}\Bigr)\biggr]^{\oh}+C
			\,,
		\end{equation}
with \eqref{LW} given by the first two terms in the large $R$ expansion. But it doesn't help a lot, as seen from the black curves displayed in Fig.3.

In short, there does seem to be evidence that in the case of the $\Sigma$ potentials the energy gaps above the ground state are finite for long 
distances, up to $2.5\,\text{fm}$. 

We must ask what will happen if $R$ exceeds $2.5\,\text{fm}$? In \cite{kuti}, it is stated that for $R>2\,\text{fm}$ the energy levels nearly reproduce the asymptotic $\pi/R$ gaps. We think that it is still an open question. The last point on the right of the plots shown in Fig.3 may be a 
lattice artefact because of large errors at $R=3\,\text{fm}$. Obviously, in the absence of these points the statement is far from being indisputable. It 
is unfortunate that at present no new data on the $\Sigma$'s exists. 

Recent investigations of \cite{teper} have considered closed flux tubes of length $R$ in $SU(N)$ lattice gauge theories.\footnote{We thank 
M. Teper for letting us know this.} In this context, there is an interesting observation that should be mentioned in relation to the above discussion. Some states with $J^P=0^-$ quantum numbers show large deviations 
from the Nambu-Goto closed string spectrum. The ansatz for the energy levels of these states is given by

\begin{equation}\label{teper}
	E_q(R)=E_0(R)+\biggl[m^2+\Bigl(\frac{2\pi q}{R}\Bigr)^2\biggr]^{\oh}
	\,,
\end{equation}
where $E_0$ is a ground state energy and $q=\{0,1\}$. The value of $m$ is fitted to $m=1.85\sqrt{\sigma}$. Note that the second term is nothing but the energy of a free particle on a circle. 

Like in the case of the $\Sigma$ potentials, the energy gaps in \eqref{teper} are finite for large $R$. Explicitly, $\Delta E_q=m+O(1/R^2)$. But it is not the whole story. It becomes more interesting as we compare the estimates for $\Delta_3$ from \eqref{fit} to $m$ from \eqref{teper}. They are given by $\Delta_3=0.77\,\text{GeV}$ and $m=0.78\,\text{GeV}$.\footnote{In doing so, we use $\sigma=0.18\,\text{GeV}^2$.} This immediately raises a question whether the same object of gluonic nature occurred in both the cases. However, the answer might be negative because of the leading order corrections given by $1/R$ and $1/R^2$ terms, respectively. 

In our discussion so far, we have not said anything about the $\Sigma$ potentials at short distances. Here our models may be inappropriate because of the assumption of locality. Nevertheless, it is worth doing so to see how everything looks and works. 

For large $\kappa$, in both models the potential after extrapolation to small $R$ behaves as 

\begin{equation}\label{Es}
	\tilde V(R)=\tilde V_0+A R^2+O(R^3)
	\,,
\end{equation}
with $\tilde V_0$ given by \eqref{tV}, \eqref{functionB0} and $A$ given by \eqref{coefA}. This is a very satisfactory result. It is known from the effective string theories in flat space \cite{arvis,kal}, the MIT bag model \cite{bag} and lattice simulations \cite{bali-pineda}.\footnote{For more discussion, see \cite{bali}.} The constant term $\tilde V_0$ is 
scheme dependent. It includes the common renormalization constant $C$, such as that in the energy of the ground state \eqref{sr}. This makes 
a quantitative comparison with others difficult. On the other hand, $A$ is scheme independent, so it will be interesting to test it.

We use the values for the parameters of Model B as those in Fig.2 to estimate the coefficients in \eqref{Es}. Then our prediction for $\Sigma_u^-$ is 

\begin{equation}\label{estimateV0}
	\tilde V_0\approx1.90\,\text{GeV}\,,\quad
	A\approx0.03\,\text{GeV}^3
	\,.
\end{equation}
Interestingly, the estimate for $\tilde V_0$ is in the range of the known estimates for the lightest gluelump mass (for discussion, see \cite{bali-pineda}).

\section{Concluding Comments}
\renewcommand{\theequation}{4.\arabic{equation}}
\setcounter{equation}{0}

(i) Although for the $\Sigma$ states the results of Sec.II hold in more general circumstances, in Sec.III A we have restricted ourselves to $\Sigma_u^-$, 
the first excited state. The second is $\Sigma_g^-$. Our results for this case are shown in Fig.4. Just as before, 
\begin{figure}[htbp]
\centering
\includegraphics[width=5.25cm]{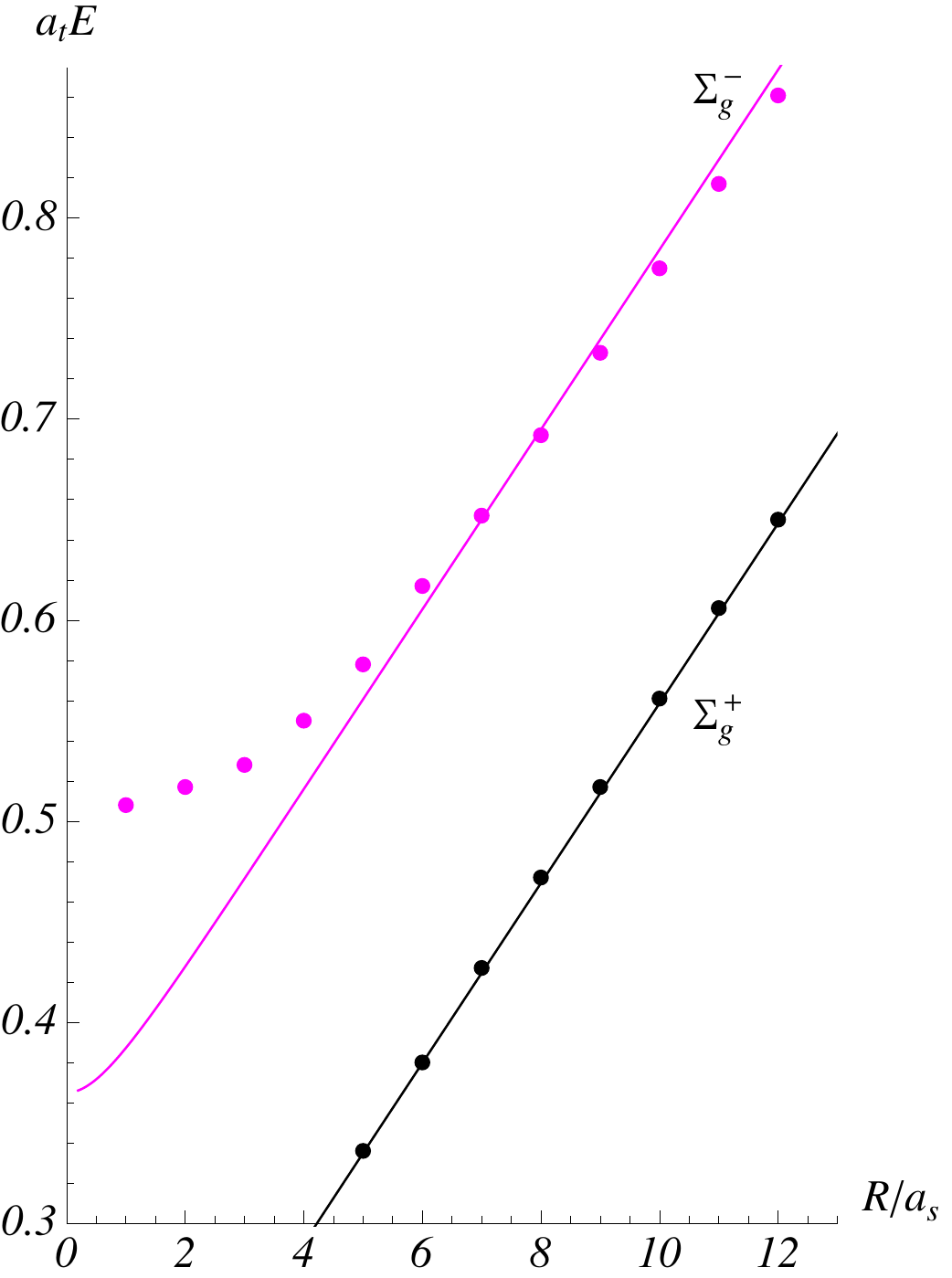}
\hspace{5cm}
\includegraphics[width=5.25cm]{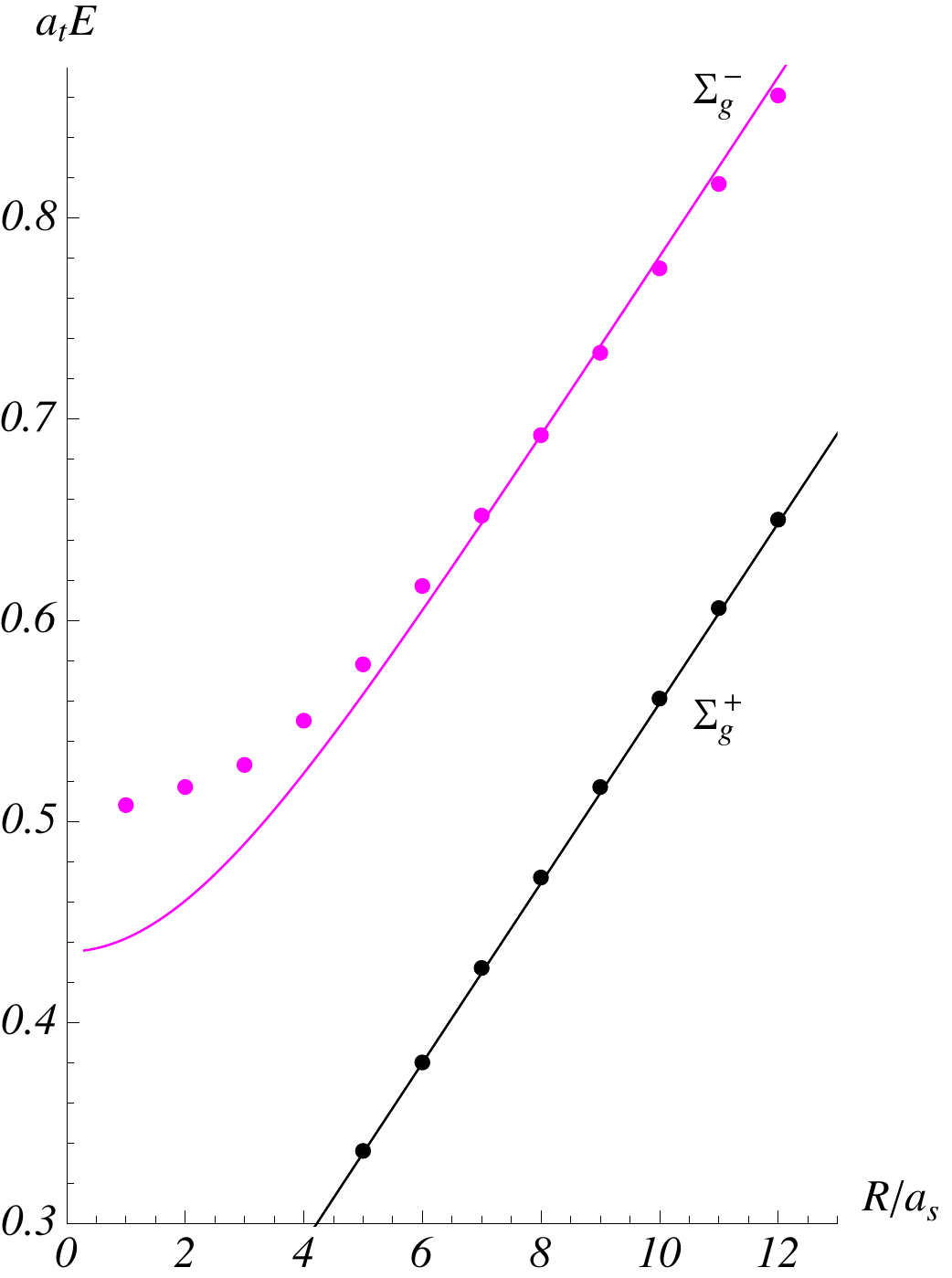}
\caption{{\small Energy levels $\Sigma_g^+$ and $\Sigma_g^-$. The dots represent the lattice data. The curves are our predictions. Left: Model A.
The value of $\kappa$ is set to $2.8$. Right: Model B. The value of $\kappa$ is set to $3800$. In both the cases we use the data and notations from Fig.2 of \cite{kuti}. So, we have $a_s=0.2\,\text{fm}$ and $a_t\approx 0.041\,\text{fm}$.}}
\end{figure}
the values of $\g$, $\s$, and $C$ are fitted to the data of \cite{kuti} for $\Sigma_g^+$. We see that for $R$ below about $0.6\,\text{fm}$ the lattice 
data are well described by neither Model A nor Model B. Again, Model B performs better but still leaves room for refinement. A possible way to do so is to consider a couple of defects as it follows from the values of $\kappa$.

The observed agreement with the lattice data for $\Sigma_u^-$ is somewhat surprising and in fact difficult to understand. Right now we don't 
have any microscopic description of the defect. So, we know nothing about its quantum numbers $(+,-)$ as well as $(g,u)$. This poses the question of 
how reliable our models at short distances are.

(ii) There are two important distinctions between static strings connecting the quark sources on the boundary of the deformed AdS space, see Fig.5. 

First, in 
\begin{figure}[htbp]
\centering
\includegraphics[width=6cm]{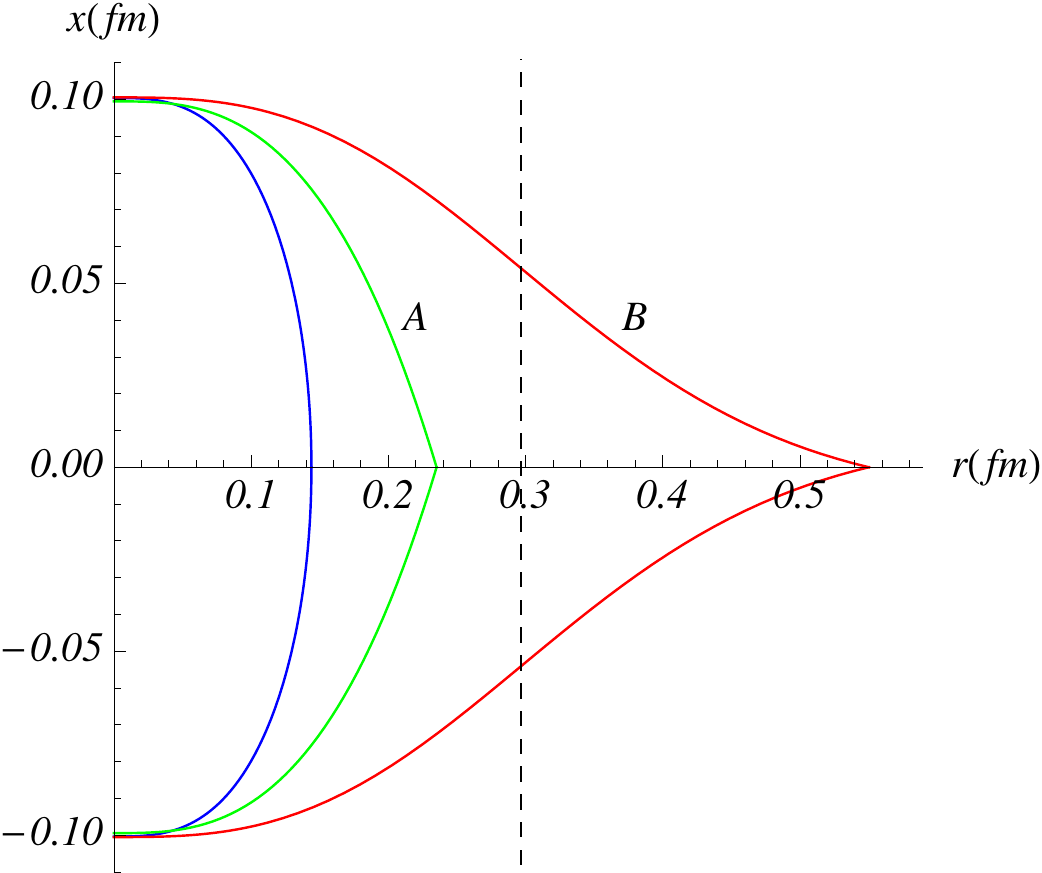}
\caption{{\small Static strings bend away from the boundary into the bulk. For models A and B, the defects are at the cusps. The quark and antiquark are set at $x=\pm 0.10\,\text{fm}$. The dashed line indicates a soft wall. 
}}
\end{figure}
Models A and B the defects lead to the formation of cusps at $x=0$. The cusp (deviation) angle is given by 
\begin{equation}\label{cusp}
\cos\frac{\theta(\lambda)}{2}=
\begin{cases}
	\sqrt{\rho}\\
	\sqrt{\bar\rho}
\,\,,
\end{cases}
\end{equation}
where the upper expression holds in Model A and the lower expression holds in Model B. By using \eqref{l-eqs-ex} and \eqref{l-eqs-ex10}, one can then 
write $\theta (R)$ in parametric form. In Model A the cusp angle becomes zero in the limit $R\rightarrow\infty$, while the effective potential \eqref{potential} remains finite. 

Interestingly, there has been an extensive discussion of cusps in the context of cosmic strings \cite{cosmicstrings}. In this case cusps can be formed 
when small-scale structures (kinks) propagating on the string in opposite directions collide with each other.

Second, in Model B the defect is allowed to go beyond the soft wall. This might have interesting implications for evaluating generalized 
Wilson loops at finite temperature. 

(iii) The parameterization proposed in this work can be obtained from that in \cite{LW} by a simple renormalization of 
the parameters. The transformation

\begin{equation}\label{renorm}
	C\rightarrow C+\Delta_n\,,\quad n\rightarrow Z_n n
	\,,\quad\text{with}\quad \Delta_0=0\,,
\end{equation}
transforms \eqref{LW} to \eqref{fit}.

It is interesting to see how this can be modeled within effective string theories in a four-dimensional flat space. A good starting point 
is to take a string ended on two heavy quarks and then load it with point-like masses.\footnote{For more discussion and references, see \cite{bn-book}.} On the other hand, it is possible to go along the way of section 2. 

For example, consider two heavy quarks separated by a distance $R$ such that each is the endpoint of an open string. The strings join at a point-like defect of mass $m$. In the static gauge the leading terms of the derivative expansion of low-energy effective action are now 

\begin{equation}\label{4dNG}
	S=-\sigma R\int d\tau-m\int d\tau+\dots
	\,,
\end{equation}
where the dots denote higher order terms. Obviously, the second term leads to a finite energy gap above the ground state at long distances $\Delta E=m+O(1/R)$ that is equivalent to the first transformation in \eqref{renorm}. The challenge is to see whether two-derivative terms in \eqref{4dNG} lead to the second transformation.

\begin{acknowledgments}
We would like to thank J. Powell for discussions and collaboration during the initial stages of this work. We also wish to thank 
J. Polchinski, E. Shuryak, M. Teper, and especially P. Weisz for helpful discussions and comments. Finally, we would like to thank  
the Institute for Nuclear Theory at UW, the Kavli Institute for Theoretical Physics at UCSB, and the Arnold 
Sommerfeld Center for Theoretical Physics at LMU for the warm hospitality. This research was supported in part by NSF 
grant PHY11-25915. 
\end{acknowledgments}

\end{document}